\documentclass[aps,twocolumn,superscriptaddress]{revtex4-1}

\usepackage{subfigure}
\usepackage{braket}
\usepackage{graphicx}% Include figure files
\usepackage{amsmath}

\begin{document}

\title{Measurement-device-independent quantum key distribution: from idea towards application}

\author{Raju Valivarthi$^{\rm ab}$, Itzel Lucio-Martinez$^{\rm ab}$, Philip Chan$^{\rm ac}$, Allison Rubenok$^{\rm abf}$, Caleb John$^{\rm ac}$, Daniel Korchinski$^{\rm ab}$, Cooper Duffin$^{\rm ab}$, Francesco Marsili$^{\rm e}$, Varun Verma$^{\rm d}$, Mathew D. Shaw$^{\rm e}$, Jeffrey A. Stern$^{\rm e}$, Sae Woo Nam$^{\rm d}$, Daniel Oblak$^{\rm ab}$, Qiang Zhou$^{\rm ab}$, Joshua A. Slater$^{\rm abg}$ {\em \&} Wolfgang Tittel$^{\rm ab}$.
\\\vspace{6pt} 
$^{a}$  {\em{Institute for Quantum Science and Technology, University of Calgary, Calgary, AB, T2N 1N4, Canada}}.\\ 
 $^{b}${\em{Department of Physics and Astronomy, University of Calgary, Calgary, AB, T2N 1N4, Canada}}.\\
$^{c}${\em{ Department of Electrical and Computer Engineering, University of Calgary, Calgary, AB, T2N 1N4, Canada}}.\\
$^{d}${\em{National Institute of Standards and Technology, Boulder, CO, 80305, USA}}.\\
$^{e}${\em{Jet Propulsion Laboratory, California Institute of Technology, Pasadena, CA, 91109, USA}}.\\
$^{f}${\em{Present Address: School of Physics, HH Wills Physics Laboratory, University of Bristol, Tyndall Avenue, Bristol, BS8 1TL, United Kingdom}}.\\
$^{g}${\em{Present Address: Vienna Center for Quantum Science and Technology (VCQ), Faculty of Physics, University of Vienna, A-1090, Vienna, Austria}}.\\}

\begin{abstract}
We assess the overall performance of our quantum key distribution (QKD) system implementing the measurement-device-independent (MDI) protocol using components with varying capabilities such as different single photon detectors and qubit preparation hardware. We experimentally show that superconducting nanowire single photon detectors allow QKD over a channel featuring 60 dB loss, and QKD with more than 600 bits of secret key per second (not considering finite key effects) over a 16 dB loss channel. This corresponds to 300 km and 80 km of standard telecommunication fiber, respectively. We also demonstrate that the integration of our QKD system into FPGA-based hardware (instead of state-of-the-art arbitrary waveform generators) does not impact on its performance. Our investigation allows us to acquire an improved understanding of the trade-offs between complexity, cost and system performance, which is required for future customization of MDI-QKD. Given that our system can be operated outside the laboratory over deployed fiber, we conclude that MDI-QKD is a promising approach to information-theoretic secure key distribution.

\end{abstract}

\maketitle

\section{Introduction}

Quantum key distribution (QKD)~\cite{BB84,Gisin2002,Scarani2009,Lo2014} is the most mature application of quantum information processing -- it allows two parties (commonly known as Alice and Bob) to exchange information-theoretic secure cryptographic keys. During the past decade, experimental work has focussed on the development of systems delivering secret keys over deployed fiber at high rates~\cite{Stucki2009,Dixon2010,Sasaki2011}, over hundreds of kilometres~\cite{Stucki2009}, and in trusted-node networks~\cite{Sasaki2011,Shields2013}. Furthermore, commercial systems are available for purchase~\cite{commercial}.

From a theoretical point of view, the security of QKD is guaranteed by the laws of quantum mechanics. However, in practice, the physical devices employed in QKD systems never perfectly agree with theoretical descriptions. This disparity has led to attacks that compromise the security of real systems~\cite{Brassard2000,Gisin2006,Makarov2006,LamasLinares2007,Fung2007,Qi2007,Zhao2008,Lydersen2010,Jain2011,Dusek2000,Tang2013,Bugge2014}. Most importantly, certain attacks~\cite{Lydersen2010,Bugge2014} known as `blinding attacks', exploit vulnerabilities of single-photon detectors (SPDs) to allow an eavesdropper to remotely control all SPDs. In fact, the majority of successful attacks have targeted the SPDs of QKD systems.

Protecting practical QKD systems against these `side-channel' attacks is a difficult problem, which is currently being investigated by many research groups. Some strategies include developing counter-measures to specific attacks~\cite{Zhao2008,Yuan2010} and developing techniques to eliminate specific weaknesses of devices~\cite{Lim2014}. However with such approaches it is difficult to fully characterize and quantify all possible weaknesses and corresponding attacks, and it may be possible for an eavesdropper to circumvent the countermeasure. Another approach is to develop protocols that are intrinsically secure against side-channel attacks, such as device-independent QKD (DI-QKD)~\cite{Acin2007,Masanes2011}, in which no devices have to be trusted and security is guaranteed via a loophole-free Bell-inequality violation. Unfortunately, despite much effort~\cite{Giustina2013,Christensen2013}, a loophole-free Bell test has yet to be achieved and the implementation of DI-QKD therefore seems unlikely in near future.
\begin{figure*}[t]
\begin{center} 
\includegraphics{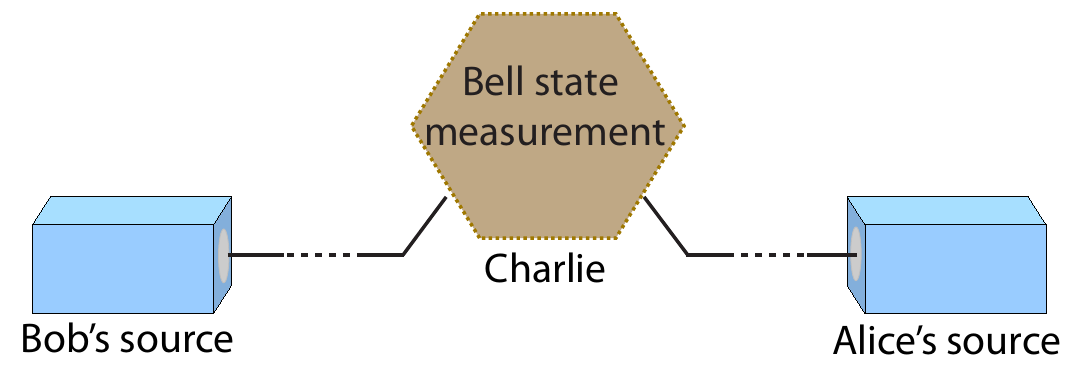} 
\caption{Schematic of the setup to implement the MDI-QKD protocol. The figure shows two qubit sources labeled as Bob's source and Alice's source that send qubits through a quantum channel to Charlie, a third untrusted party, who performs a Bell state measurement.} 
\label{scheme} 
\end{center}
\end{figure*}
\newline \indent More recently, several groups have proposed QKD protocols that are intrinsically secure against the most dangerous side-channel attacks, namely all possible (i.e. known or yet-to-be-proposed) detector side-channel attacks, while requiring proper functioning of other devices~\cite{Lo2012,Braunstein2012,Gonzalez2014,Cao2014,Geneva2014}. The first such protocol, known as measurement-device-independent QKD (MDI-QKD)~\cite{Lo2012}, was inspired by time-reversed entanglement-based QKD~\cite{Biham1996,Inamori2002} and, for maximum secret key rate, requires a Bell state measurement (BSM) at a central station (henceforth referred to as Charlie) to create entanglement-like correlations between Alice and Bob. Note that even if an adversary completely controls the measurement device (including the detectors) and, e.g. replaces the entangling measurement by any other measurement, he would not gain any information about the cryptographic key. Hence, as no assumptions about the functioning of the measurement apparatus are required, MDI-QKD is intrinsically immune to all detector side-channel attacks. Thus, MDI-QKD provides enhanced security as compared to traditional QKD.

The MDI-QKD protocol has other important benefits. For instance, it is natural to extend the MDI-QKD scheme to star-type network topologies, in which a large number of users is connected to the same central station, i.e. Charlie, who connects pairs of users on demand. Such a design is more cost-effective per user than quantum networks comprised of individual point-to-point links~\cite{Shields2013}. Also, MDI-QKD is a natural stepping stone towards quantum repeaters, which is one approach towards truly long-distance quantum communication~\cite{Sangouard2011}.

On the other hand, similar to the case of quantum repeater-based communication, MDI-QKD faces the challenge of BSMs with photons created by distant, and independent, photonic qubit sources. Such a measurement requires nearly complete indistinguishability of the photons from each source upon arrival at Charlie's, including polarization, temporal and spectral profiles. The latter is determined largely by local properties of the sources and can thus be easily controlled. However, polarization and arrival-time fluctuate due to time-varying properties of the entire quantum channel between the sources and the central station, as birefringence and refractive index of optical fiber are temperature dependent. Ensuring indistinguishability thus requires feedback systems that counteract external environmental changes. The first demonstration of a BSM with photons from independent and widely separated sources was demonstrated only recently~\cite{Rubenok2013} -- explicitly for MDI-QKD.

Due to its many advantages, MDI-QKD has received much attention from experimental groups and has been demonstrated in several configurations. The feasibility of MDI-QKD was first demonstrated in~\cite{Rubenok2013} using time-bin encoding over up to 80 km of spooled fiber as well as over 18.6 km of deployed fiber across a city centre, and in~\cite{Liu2013} (using the same type of encoding) over 50~km of spooled fiber and with random modulation of all bases and states, as required for actual key distribution. Subsequent experiments employed polarization qubits in a lab setting with quantum signals frequency-multiplexed with classical signals ~\cite{daSilva2013}, and with pre-set random state and basis modulation \cite{Tang2014}. Most recently, MDI-QKD has been demonstrated over 200 km of spooled optical fiber~\cite{Tang200}, and in fully-automated fashion and over a real-world link~\cite{Tang30}.

Our previous experiments have demonstrated the proof-of-principle of MDI-QKD over spooled as well as deployed fiber using a particular configuration for Alice, Bob and Charlie \cite{Rubenok2013} (in these measurements, only the quantum channel changed). Here we assess the impact of using  components with varying capabilities -- single photon detectors and qubit-preparation hardware -- on overall system performance, i.e. secret key rates, and maximum tolerable transmission loss. This allows us to develop a better understanding of the trade-offs between complexity, cost, and system performance, which is required for future customization of QKD systems.

This paper is organized as follows: In section~\ref{protocol}, we describe the MDI-QKD protocol and in section~\ref{sec:realization}, a detailed description of the realization of our MDI-QKD system  is presented. In section~\ref{result}, we specify different configurations in which our MDI-QKD system is tested and then discuss results of these tests. We conclude in section~\ref{conclusion}.

\section{The MDI-QKD protocol}
\label{protocol}

Sources of quantum states are at Alice's and Bob's stations. To encode classical bits, they randomly choose a basis and a state among the four BB84 states. In the case of time-bin qubits, these may correspond to the states $\ket{e}$ and $\ket{l}$ (i.e. eigenstates of the Pauli operator $\sigma_Z$, forming the so-called Z-basis), or to $\ket{+} \equiv (\ket{e} + \ket{l})/\sqrt{2}$ and $\ket{-} \equiv (\ket{e} - \ket{l})/\sqrt{2}$ (i.e. eigenstates of the Pauli operator $\sigma_X$, forming the so-called X-basis), where $\ket{e}$ and $\ket{l}$ denote the emission of a photon in an early or late temporal mode, respectively. Furthermore, they associate $\ket{e}$ and $\ket{+}$ with a classical bit value of 0, and $\ket{l}$ and $\ket{-}$ with a classical bit value of 1. The qubits are sent through quantum channels to a third station at which Charlie performs a Bell state measurement (see Figure \ref{scheme}) that projects their joint state onto one of the four maximally entangled Bell states: 

\begin{eqnarray}
\ket{\psi^\pm_{AB}} =\frac{1}{\sqrt{2}}( \ket{e_A l_B} \pm  \ket{l_A e_B}), \nonumber \\
\ket{\phi^\pm_{AB}} =\frac{1}{\sqrt{2}}( \ket{e_A e_B} \pm  \ket{l_A l_B}).
\label{BSM}
\end{eqnarray}

Once a sufficient number of qubits has been transmitted \cite{Curty2014}, Charlie announces which of his joint measurements resulted in one of the four states of Eq.~\ref{BSM}, as well as the result of the measurement (note that to ensure security, Charlie only needs to be able to project onto one Bell state, but access to more Bell states will increase performance). As a next step, Alice and Bob perform a basis reconciliation procedure known as key sifting. In this step, for each successful projection onto a Bell state, Alice and Bob reveal and compare the bases employed to prepare their respective qubits over an authenticated classical channel and keep only the events in which they used the same basis. Next, depending on the result of the BSM and his preparation basis, Bob must post-process his bit values to ensure identical bit values to Alice. For example, Bob performs a bit flip if the announced measurement resulted in $\ket{\psi^-_{AB}}$, which indicates anti-correlated bits in both the X- as well as the Z-basis. Furthermore, due to unavoidable experimental imperfections such as faulty qubit preparation, channel noise and noisy single photon detectors, as well as possible eavesdropping, it is necessary to go through a key distilling process: After publicly revealing a subset of their prepared bit values, Alice and Bob estimate the error rate for each basis independently, and then perform error correction on the Z-basis bits to remove all discrepancies in their Z-keys (i.e. the key bits associated with preparations in the Z-basis). The error rate for the X-basis is used to bound information that an eavesdropper could have obtained during photon transmission and detection; it is removed by means of privacy amplification. The final secret key rate is given by: 
\begin{equation}
S \geq [Q^z[1-h_2(e^x)] - Q^zf h_2(e^z)]
\label{keyrate_basic}
\end{equation}
in which $Q$ refers to the gain (the probability of a projection onto a Bell state per emitted pair of qubits), $e$ indicates error rates (the ratio of erroneous to total projections a Bell state), $h_2$ is the binary Shannon entropy and $f$ refers to the efficiency of error correction with respect to Shannon's noisy coding theorem. The superscripts $x$ or $z$ indicate to which basis a particular variable refers.

Due to the need for qubits encoded into individual photons, the above-described implementation is currently difficult to implement. However, by taking advantage of so-called decoy states, it is possible to use phase-randomized weak coherent states (i.e. attenuated laser pulses, which sometimes contain more than one photon) ~\cite{Lo2012,Wang2013,Ma2012,Curty2014}, which are straightforward to generate with current technology. This is similar to traditional prepare \& measure-type protocols~\cite{Hwang2003,Lodecoy,Wang2005}. By randomly modulating the mean photon number of the laser pulses between several values known as `signal' and `decoy' (the optimal values of which depend on the implementation \cite{Chan13,Xu2014}), one can assess the gains and error rates associated with single-photon emissions. The secret key rate is then given by:
\begin{equation}
S \geq [Q_{11}^z[1-h_2(e^x_{11})] - Q^z_{\mu\sigma}f h_2(e^z_{\mu\sigma})].
\label{keyrate}
\end{equation}
The `$11$' subscript indicates values for gain and error rate stemming from Alice and Bob both emitting a single photon, and the subscript `$\mu\sigma$' denotes values associated with mean photon numbers per `signal' pulse of $\mu$ and $\sigma$, emitted by Alice's and Bob's, respectively. As in Eq. \ref{keyrate_basic}, the basis is indicated using a superscript.

\section{Realization of our MDI-QKD system}

The experimental setup of our MDI-QKD system can be divided into four parts: qubit preparation, quantum channel, feedback systems, and the BSM unit. We have tested our system with various sources and single photon detectors, and in the following subsections we present a detailed description of each part of our system for each of the configurations used in our MDI-QKD demonstrations.
\label{sec:realization}

\begin{figure*}[ht]
\begin{center} 
\includegraphics{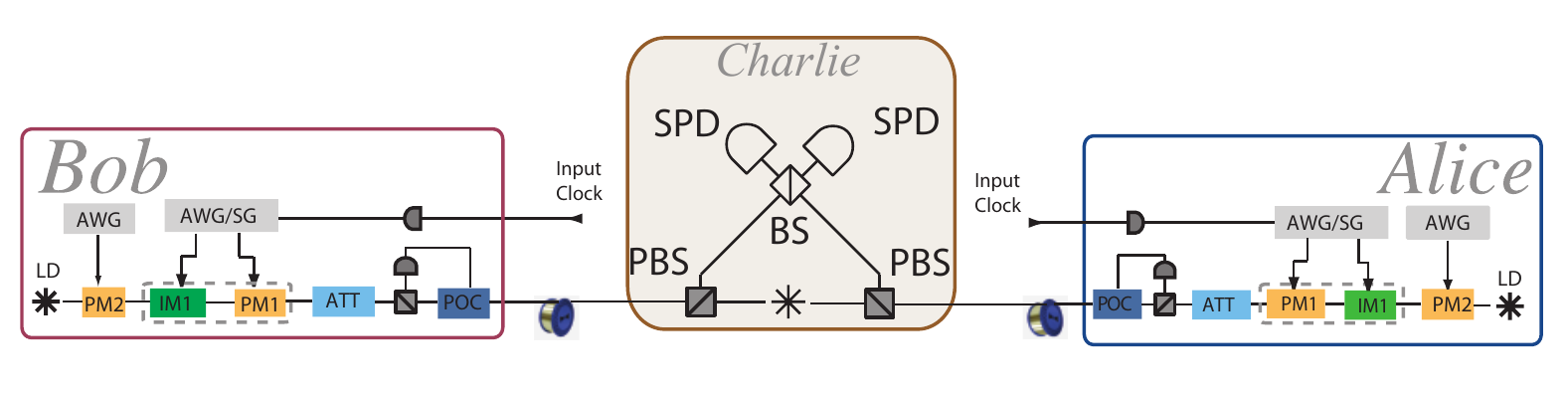} 
\caption{Experimental setup of our MDI-QKD system (see main text for details). LD: laser diode, PM1 and PM2: phase modulator, IM: intensity modulator, ATT: attenuator, POC: polarization controller and measurement, AWG/SG: arbitrary waveform generator or FPGA-based signal generator (depending on implementation), PBS: polarization beam splitter, BS: beam splitter, SPD: single photon detector.}
\label{setup} 
\end{center}
\end{figure*}

\subsection{Qubit preparation}
\label{qubitprep}
In our MDI-QKD system, Alice and Bob prepare time-bin qubits in one of the four BB84 states introduced above. In our work, time-bin qubits are prepared by externally modulating a continuous wave (CW) laser at 1552 nm wavelength with a commercial intensity modulator (IM1) and a phase modulator (PM1), as depicted in Figure~\ref{setup}. The time-bin qubits in the X- and  Z-basis are prepared by providing different electronic pulses to IM1 and PM1. By changing the attenuation of the optical attenuator (ATT), different mean photon numbers for the time-bin qubits are obtained, as required for the decoy state protocol. In our experiments, the mean photon numbers of the signal and decoy states are optimized through a theoretical model of MDI-QKD  to obtain the highest final key rate for each distance~\cite{Chan13}. The phase of the qubits is uniformly randomized by applying an electronic signal with randomized amplitudes to PM2, which ensures protection against the unambiguous-state-discrimination (USD) attack ~\cite{Dusek2000,Tang2013}. Note that PM2 was not included in some of the measurements described below, and some measurements furthermore employed only one CW laser, whose output was split and sent to Alice and Bob for qubit preparation (deviations from the generic setup depicted in Figure \ref{setup} will be mentioned again in section \ref{result}). It is also worth noting that time-bin qubits can also be prepared using a directly modulated pulsed laser in conjunction with an unbalanced Mach-Zehnder interferometer, IMs and PMs ~\cite{Liu2013}.

For our experiments, we use and compare two different electronic driving devices: a commercially-available arbitrary waveform generator (AWG) and a home-made signal generator (SG) based on a field programmable gate array (FPGA) and suitable drivers that transform the FPGA outputs into appropriate analog signals. We developed our signal generator as a cost-effective and compact solution to replace the AWG in anticipation of future development of commercial MDI-QKD system. With the AWG we generate time-bin qubits with a temporal mode width of 100 to 500 ps and a temporal mode separation between 1.5 and 2.5 ns. With our signal generator we generate time-bin qubits with a temporal mode width of 290 ps and a temporal mode separation of 2.5 ns.
We characterize the time-bin qubits according to the procedure described in~\cite{Chan13}. Motivated by our setups, which create imperfect pure states, we consider time-bin qubits of the form

\begin{multline}
\ket{\psi} = \frac{1}{\sqrt{1+2b^{x,z}}}\big(\sqrt{m^{x,z}+b^{x,z}} \ket{e} \\
+ e^{i\phi^{x,z}}\sqrt{1-m^{x,z}+b^{x,z}} \ket{l}\big).
\label{eqn:photon_state}
\end{multline}

Here $\ket{e}$ and $\ket{l}$ denote orthogonal early and late temporal modes, respectively. $m^{x,z}$ and $b^{x,z}$ are determined by the properties of the electronic signals from the AWG/SG (e.g. the amplitude, and rising and falling edges of the pulses), and the extinction ratio (ER) of IM1. Furthermore, the superscript indicates to which basis/state the parameter applies (e.g. $m^{z=0}$ applies to the state produces when Alice or Bob chooses the Z-basis state $\ket{l}$).  For a perfect time-bin qubit preparation setup, $m^{z=0,1}$ = 0 or 1 for qubits in the Z-basis, and $m^{x=+,-}$ = 0.5 for qubits in the X-basis; $\phi^{x=+,-}$ = 0 or $\pi$ for qubits in the X-basis (in the Z-basis $\phi^{z}$ is irrelevant) and $b^{x,z}$ would be zero. We measured these parameters for both implementations; the results are given in Table~\ref{Qubits_AWG}.
As one can see, the parameters barely change when moving from an AWG to our signal generator.  The impact on overall system performance will be described in section~\ref{qpr}.

\begin{table}[h!!!!!!!!!!!!!!]
\begin{tabular}{l l l}
\hline
 Parameter & AWG & SG \\
\hline
  $b^{z=0,1,x=-,+}$  & (5.34~$\pm$~0.91)$\times$$10^{-5}$ & (2.57~$\pm$~1.18)$\times$$10^{-5}$ \\
  $m^{z=0}$  & (0.986~$\pm$~0.012) & (0.982~$\pm$~0.010) \\
  $m^{z=1}$  & 0    & 0 \\
  $m^{x=-,+}$  & (0.4963~$\pm$~0.0042) & (0.4963~$\pm$~0.0062) \\
  $\phi^{z=0,1} $=$~\phi^{x=+}[rad] \ \ $&0&0\\
  $\phi^{x=-}[rad] $&$\pi$+(0.075~$\pm$~0.015) \ \ &$\pi$+(0.075$\pm$0.015)\\
\hline
\end{tabular}
\caption{Measured values for $m^{x,z}$ and $b^{x,z}$ for time-bin qubits prepared using an arbitrary waveform generator (AWG) and our FPGA-based signal generator (SG), respectively.}
\label{qubitcharac}
\label{Qubits_AWG}
\end{table}

\begin{figure}[b] 
\begin{center}
\includegraphics[scale=0.55]{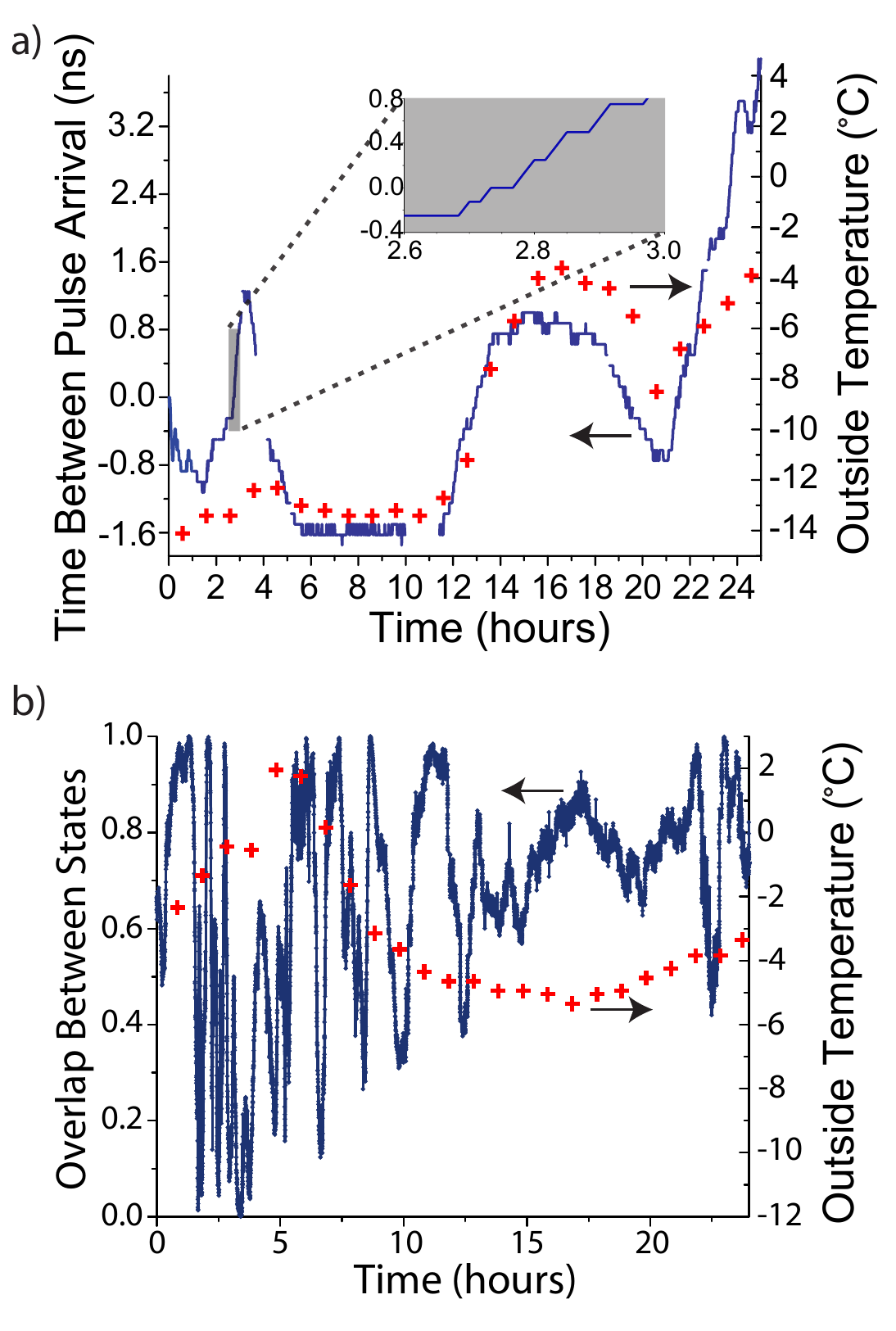} 
\caption{(a) Differential arrival time of attenuated laser pulses propagation from Alice/Bob to Charlie. (b) Variation in polarization overlap of originally horizontally polarized light propagation from Alice/Bob to Charlie. The temperature data (crosses) is given in the secondary y-axes, showing correlation with variations of differential arrival time and polarization overlap (the figure is identical to that in~\cite{Rubenok2013}).} 
\label{PRL_result} 
\end{center}
\end{figure}

\subsection{Quantum channel}
After preparing the time-bin qubits, Alice and Bob send them to Charlie through two different quantum channels (i.e. optical fibers). First, we note that loss during transmission (in conjunction with detector noise) increases the error rate, and this limits the maximum distance for MDI-QKD.
Second, ideally, the propagation times of the photons through the fibers should remain constant, and the polarization state of the time-bin qubits should not be affected. Unfortunately, both properties change due to dynamic properties of real-world fiber links. Figure~\ref{PRL_result} (a) shows the change of the differential arrival time of attenuated laser pulses from Alice's and Bob's, which we previously reported in~\cite{Rubenok2013}, over a deployed fiber across the city of Calgary; Figure~\ref{PRL_result} (b) depicts the overlap between the polarization states of originally horizontally polarized light from Alice and Bob after propagation to Charlie. It can be seen that both the differential arrival time and the overlap of the polarization states vary over time and are correlated to the outside temperature. As shown in Figure~\ref{PRL_result}, the differential arrival time varies from -1.6 ns to 0.8 ns in three hours (hours 12--15), during which the temperature increases from -14 $^{\circ}$C to -4 $^{\circ}$C; and the polarization overlap can vary from 0 (orthogonally polarized pulses) to 1 (identically polarized pulses) in much less than one hour. Hence, to realize a stable MDI-QKD system, we developed active feedback subsystems that control the propagation time and polarization fluctuation, the details of which are given in section \ref{sec:feedback}.

\subsection{Feedback systems}
\label{sec:feedback}
The main technological challenge to implement MDI-QKD is to perform a BSM with photons from independent sources that travelled through two independent fibers. For this measurement, one requires the two photons to be approximately indistinguishable, i.e, they should have sufficient temporal, polarization, and spectral overlap. To this effect, we implement different feedback mechanisms.
\subsubsection{Temporal overlap}
Charlie distributes optical clock signals (at 10 MHz) to Alice and Bob via a second optical fiber, which they convert to electrical pulses. This master clock is used to synchronize all relevant devices at Alice's, Bob's and Charlie's. To ensure sufficient temporal overlap, Charlie measures the arrival time of Alice's and Bob's pulses independently every few minutes using a time-to-digital converter (TDC) and SPDs. He then applies an appropriate delay to the distributed clock signal to match the arrival time difference of signals emitted at Alice's and Bob's within 30 ps.
\subsubsection{Polarization overlap}
Charlie sends strong vertically polarized light for 250 ms every 10 seconds through the links to Alice and Bob. Alice and Bob measure its polarization and prepare their qubits orthogonal to that. This ensures that the qubits from Alice and Bob, after travelling through the link, will be identically (horizontally) polarized at Charlie and arrive at the BS. Note that the PBSs in Charlie's system ensure that polarization fluctuations transform into added loss, and not decreased indistinguishability. For additional details see \cite{Rubenok2013}.
\subsubsection{Spectral overlap}
Alice and Bob split some of their laser light that is used to prepare qubits and continuously send it to Charlie via the second optical fiber. Charlie monitors the frequency difference between the two lasers using their beat note signal. Whenever the frequency difference is greater than the threshold of 10 MHz, Charlie communicates the frequency difference to Alice. To maintain sufficient spectral overlap, Alice uses a frequency shifter comprising a phase modulator to which appropriate linear phase chirps are applied using a serrodyne modulation signal.

\begin{table*}[t]
\begin{tabular}{llllllll}
\hline
Configuration \ \ & Lasers \ \ & PM2 \ \ & Qubit generation \ \ & Detector \ \ &  Channel & Total channel length \ \ & Total channel loss\\
\hline
1 &2&no& AWG & id201 & Deployed \ \ &18.6 km & 9 dB \\
2 &2&no& AWG &  id201 & Spool &20, 40, 60 km&9.1, 13.7, 18.2 dB\\
3 &1&yes& AWG &  id210 & Spool&60, 100 km& 13.7, 20 dB\\
4 &1&yes& AWG & SNSPD & ATT&-&16, 40, 60 dB\\
5 &1&yes& SG & SNSPD &ATT&-&16 dB\\
\hline
\end{tabular}
\caption{MDI-QKD system configurations assessed in this work. We list the number of lasers used, the presence of PM2, the hardware employed for qubit generation (AWG or SG), and the single photon detectors (id201, id210 or SNSPD). Quantum channels include short fibers with attenuators (ATT), spooled fiber, and deployed fiber across the city of Calgary.}
\label{configurations}
\end{table*}

\subsection{BSM unit}
The BSM unit for time-bin qubits consists of a 50:50 beam splitter followed by two SPDs, as shown in Figure~\ref{setup}. The two-photon projection measurement occurs by overlapping the two photons on the beam splitter -- to erase which-way information -- and subsequently detecting the two photons. A projection onto $\ket{\psi^{-}_{AB}}$ is characterized by a coincidence detection in orthogonal temporal modes in different detectors while a projection onto $\ket{\psi^{+}_{AB}}$ is characterized by a coincidence detection in orthogonal temporal modes in the same detector. All other coincidence detections (i.e., detections in the same temporal mode) project onto product states.  It has been shown that the Bell state measurement efficiency (i.e. the probability that two independent photons are projected onto an entangled state) is limited to 50\% when using only linear optics and no auxiliary photons~\cite{BSM50}. Note that we only monitor projections onto $\ket{\psi^-_{AB}}$ for the results discussed here, which reduces the maximum efficiency (assuming lossless detection) to 25\%. However, BSM projections onto both $\ket{\psi^{-}_{AB}}$ and $\ket{\psi^{+}_{AB}}$ have been demonstrated~\cite{Raju2014}.

\begin{table}[b]
\begin{tabular}{llll}
\hline
  SPD
  & id201 & id210 & SNSPD \\
\hline
  Dark count probability$^{\rm a} \ $ & $\sim$10$^{-4}$ &  $\sim$10$^{-5}$ & $\sim$10$^{-7}$ \\
  Deadtime  & 10$~\mu$s & 10$~\mu$s & 40~ns \\
  Maximum gate rate & 8 MHz & 100 MHz \ \ & free running\\
  Detector efficiency & $\sim$15\% &$\sim$15\% &$\sim$50\%\\
  Maximum count rate & 0.1 MHz & 0.1 MHz & 2 MHz \\
  Afterpulsing$^{\rm a}$ & $\sim$ 10$^{-5}$& $\sim$10$^{-5}$ & $\sim$ 0 \\
  Temperature & $\sim$ 223 K \ \ & $\sim$ 223 K & $\sim$1 K \\
\hline
\end{tabular}
\caption{Properties of various SPDs we employed ($^{\rm a}$per ns)}
\label{tab:prop}
\end{table} 
The performance of the BSM unit, and in turn an MDI-QKD system, is determined by several detector properties, including the detection efficiency $\eta_{det}$, noise, gate rate, recovery time, and detector type. For our measurements, we employ three different types of detectors: two different types of InGaAs-based SPDs from idQuantique (id201 and id210), as well as superconducting nanowire single photon detectors (SNSPDs)~\cite{snspds}. The technical specifications are summarized in Table~\ref{tab:prop}. While the SNSPDs have far superior properties, they are more cumbersome to use due to more stringent cooling requirements.

\section{Performance of different MDI-QKD configurations -- measurements, results and discussion}
\label{result}
We tested our system in 5 different configurations (see Table~\ref{configurations}) and over different quantum channels, which allows us to assess the trade-offs between system performance (in terms of secret key rates or maximum distance), complexity, and cost. The configurations are characterized by different qubit-generation hardware (AWG or SG), the number of lasers employed (1 or 2), the presence of PM2 (used to phase-randomize the laser pulses), and the type of SPDs (id201, id210, SNSPD). Quantum channels are either short fibers with variable loss (implemented using variable attenuators), spooled fiber (without additional attenuator), or deployed fiber. 

For each configuration, we created the same combination of quantum states and intensities (i.e. signal or decoy states) at Alice's and Bob's until we gathered enough data, and then changed the states and/or the intensities as required by the decoy state protocol described in \cite{Wang2013}. Using Eq.~\ref{keyrate}, we then calculated the secret key rates per gate and per second, which are shown in Figures~\ref{key rate} and \ref{key rate bps}, respectively (Figure~\ref{key rate} additionally shows predicted values using the model described in \cite{Chan13}). In the following sections we will discuss the impact of changing components of our QKD system on its performance.

\subsection{Quantum channels -- real-world versus spooled fibers}
Configurations 1 and 2 are identical except for the quantum channel -- real-world fiber deployed across the city of Calgary, and spooled fiber inside a temperature-controlled laboratory, respectively (these results have already been presented in~\cite{Rubenok2013}). In the absence of significant environmental effects on the deployed channel, the corresponding secret key rates would be identical. However, as described above, the differential arrival time and the polarization overlap vary on time-scales of less than a minute. Hence, without properly functioning active stabilization, secret key rates would be significantly reduced. However, as one can see in Figure~\ref{key rate} we find nearly identical secret key rates, indicating that our feedback systems compensate for the environmental fluctuations as well as time-varying laser frequency differences. Note that our feedback systems have bandwidth limitations (which we have not yet assessed), i.e. will compensate only for sufficiently slow fluctuations. This may affect the maximum distance over which our QKD system can be employed outside the laboratory, even if it still delivers secret key over spooled fiber of the same length inside the lab.

\subsection{Qubit preparation -- AWG versus FPGA-based signal generator}
\label{qpr}
Next, we compare the performance obtained with configuration 4 and 5, which only differ in the hardware used for preparing qubits --  AWGs or FPGA-based signal generators (SG). Due to bandwidth differences between the AWG and the SG, the parameters of the prepared time-bin qubits -- temporal mode width and temporal mode separation -- are insignificantly different (see Table~\ref{tab:qubits}). 
The measured secret key rates per gate for a 16 dB loss channel are shown in Figure~\ref{key rate}, along with our theoretical prediction. We find very good agreement in case of configuration 5 (SG-based qubit preparation), and reasonable agreement in case of configuration 4 (AWG-based qubit preparation). We attribute the difference mainly to statistical fluctuations (both points overlap within their 2$\sigma$ uncertainty). Hence, we find that our home-made, cost-effective signal generator produces quantum states of similar quality as the state-of-the-art AWG. This conclusion is also supported by the data in Table~\ref{qubitcharac}.
\begin{table}[h]
\begin{tabular}{lll}
\hline
 Devices & Temporal mode & Temporal mode \\
 & width (FWHM) (ps) \ \ \ & separation (ns) \\
\hline
  AWG & 250 &  2.50 \\
  FPGA board \ \ \  & 290 &  2.50 \\
\hline
\end{tabular}
\caption{Parameters of prepared time-bin qubits for configurations 4 and 5 (see Table~\ref{configurations}).}
\label{tab:qubits}
\end{table}

\subsection{Impact of different single photon detectors}
\begin{figure}[t]
\begin{center}
\includegraphics[width=\columnwidth]{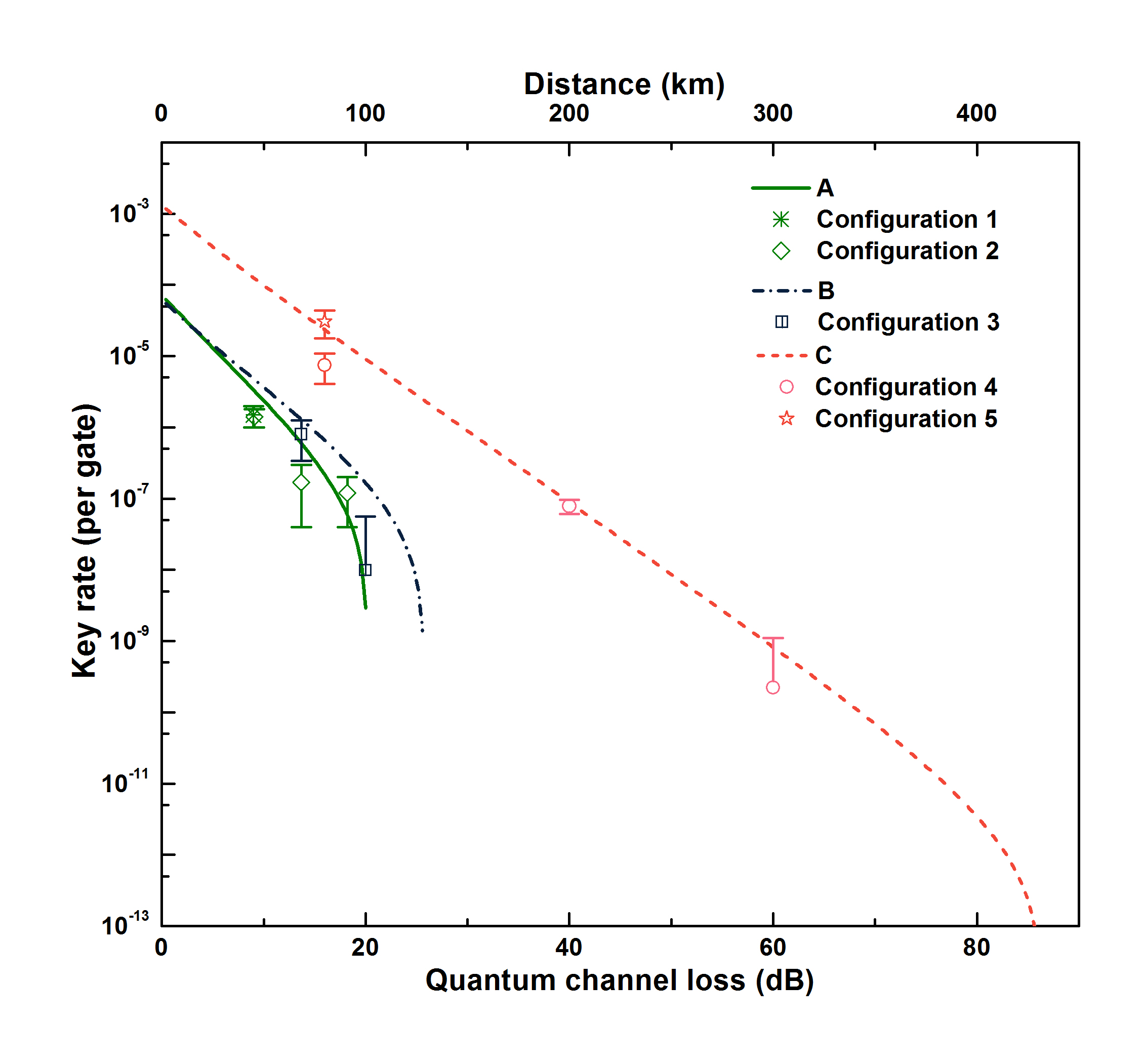}
\caption{Experimentally obtained secret key rates per gate versus total loss in the quantum channel. Curve A shows the theoretical prediction assuming configuration 1 or 2, curve B assuming configuration 3, and curve C assuming configuration 4 or 5. Experimentally obtained key rates for the different configurations are plotted as well. All configurations are described in Table~\ref{configurations}. The secondary x-axis shows distance assuming 0.2 dB loss per kilometre of fiber. Uncertainties (one standard deviation) are calculated assuming Poissonian statistics of all measured gains and error rates.}
\label{key rate}
\end{center}
\end{figure}

\begin{figure}[t]
\begin{center}
\includegraphics[width=\columnwidth]{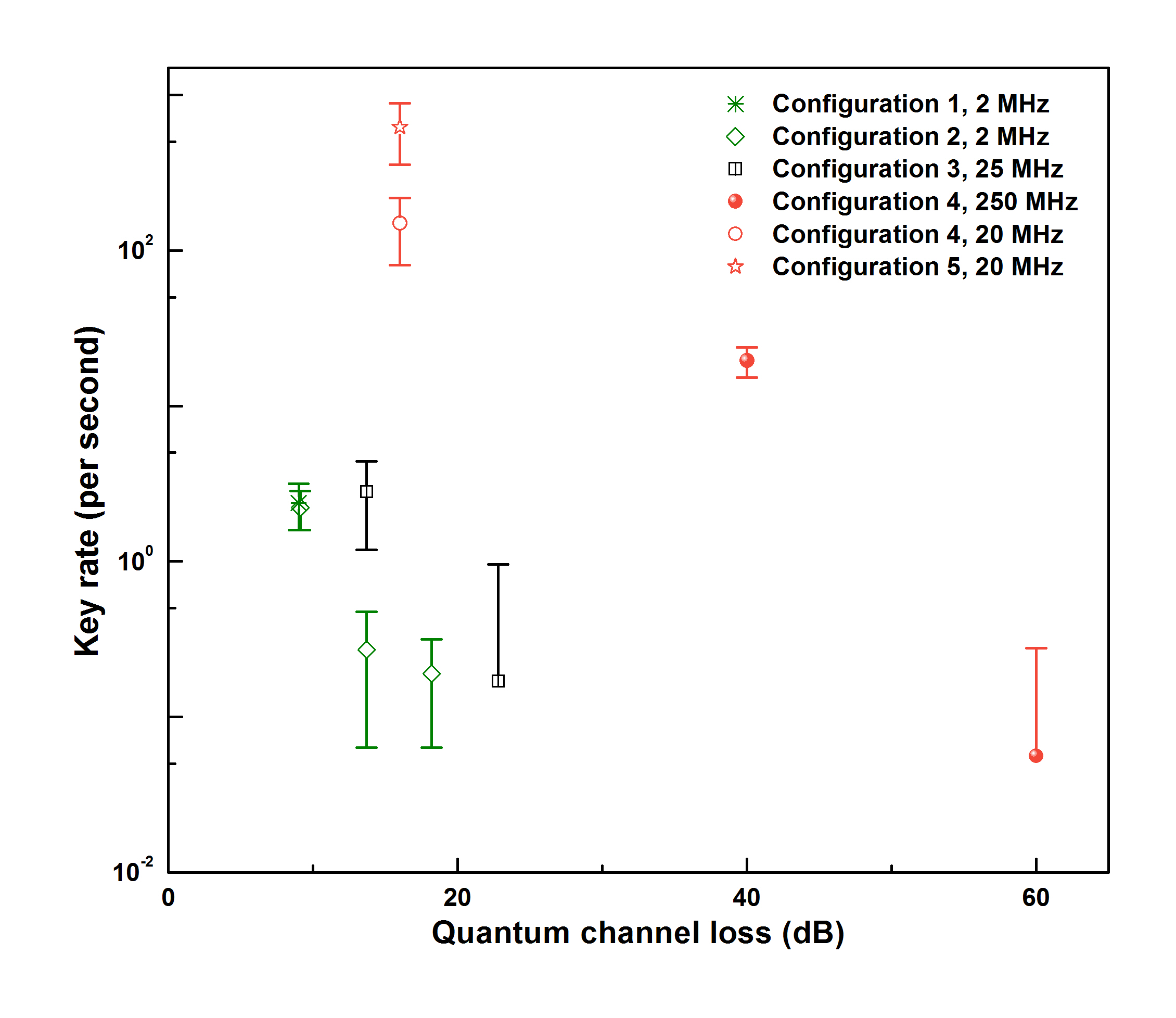}
\caption{Secret key rates per second versus loss in the quantum channel for all configurations specified in Table~\ref{configurations} and for different qubit generation rates. Uncertainties (one standard deviation) are calculated assuming Poissonian statistics of all measured gains and error rates.}
\label{key rate bps}
\end{center}
\end{figure}

Configurations 1-5 differ in the single photon detectors employed for the BSM, the quantum channel (deployed fiber, spooled fiber and variable attenuator), the number of lasers used to create qubits, as well as the hardware employed for the latter. However, given that our feedback system operates reliably (as discussed in section \ref{sec:feedback}), and ignoring the small difference between AWG and SG-based qubit creation (see section \ref{qubitprep}), we assume in the following that all differences in secret key rates have to be attributed to different SPDs. (Detector performance is characterized in Table~\ref{tab:prop}.) From the data shown in Figures~\ref{key rate} and \ref{key rate bps}, we find that the secret key rate (in bits per gate) is significantly impacted by the use of different detectors. 

As expected, because of their superior detection efficiency and lower noise, the key rate per gate for a specific amount of channel loss (i.e. distance) is higher with SNSPDs compared to id201 and id210 InGaAs SPDs. For the same reason, the maximum distance between Alice and Bob is larger. More precisely, the maximum distance over which MDI-QKD can be performed using id201 detectors is 80 km. Using id210 detectors it increases to 125 km, and for SNSPDs we find 430 km (these estimations assume fiber loss of 0.2 dB/km. Note that fibers with less loss are now available \cite{Corning}, but are probably not yet deployed). At distances below $\sim$50km, the key rate per gate when using id201 detectors is close to that of using id210 detectors because of similar detection efficiencies (solid and dash-dotted curves in Figure~\ref{key rate}), but employing id210 detectors results in greater distances because of their lower noise (refer to Table~\ref{tab:prop}).  

It is worth noting that, as shown in Figure~\ref{key rate bps}, when using SNSPDs, the secret key rate (in bits per second -- not per gate) over low-loss links is limited by the detectors' maximum count rates, in our case around 2 MHz. To ensure that we do not exceed this rate (in which case the detectors would stop operating), we reduce our qubit generation rate from its maximum of 250~MHz to 20~MHz as the quantum channel loss is reduced from 40~dB to 16~dB. Furthermore, when using InGaAs based detectors, the final key rate is always limited by the deadtime we set to minimize the afterpulsing, $\sim$ 10 $\mu$s. Figure~\ref{key rate bps} shows the secret key rate in bits per second when different detectors are employed (the figure also specifies different qubit generation rates). We find the maximum key rate to be 623 Hz (obtained with SNSPDs over a total channel loss of 16 dB), and that the key rate per second is higher using id210 detectors than with id201 detectors. This is due to the lower noise and higher gate rates (25 MHz compared to 2 MHz) of the id210 detectors (for more details see \cite{application note}). Note that finite key effects were not taken into account in calculating secret key rates \cite{Curty2014}.

\section{Conclusions and Outlook}
\label{conclusion}
In conclusion, we have tested MDI-QKD in various configurations, differing mainly in the type of detectors employed, the need for active feedback systems, and the complexity and cost of the hardware used to create time-bin qubits. We find that time-varying properties of optical fibers do not impact the system performance as they can be compensated using simple active feedback systems, and that the system can be operated using FPGA-based qubit creation hardware. We point out that, while none of our configurations is a complete QKD system, our experiments are sufficiently developed to predict the performance of a full system and understand the remaining challenges in building one. For example, there are no difficulties related to combining all features that we demonstrated above, e.g. phase randomization of weak coherent states, FPGA-based signal preparation, widely separated locations, and efficient and low noise single photon detection. Furthermore, implementing the decoy-state-protocol based on quantum random number generators (QRNGs) connected to our signal generators and drivers would, first, neither change our results in Figure~\ref{key rate} nor our conclusions, and second, is well understood~\cite{Sasaki2011}  and thus straightforward, albeit laborious, to do. Moreover, the operation of high-rate QRNGs, while a challenge, has been reported before~\cite{Gabriel10,LiuGuo2013,England14} and is independent of the choice of the QKD protocol. The same applies to real-time, fast and efficient, error correction and privacy amplification~\cite{Sasaki2011}. Hence, we do not foresee any significant challenges in creating a complete QKD system.

These findings demonstrate that MDI-QKD is already suitable for compact and cost-effective real-world implementations, even though the protocol was proposed only in late 2011. Furthermore, we find that secret key rates benefit from using SNSPDs. The drawback of being more expensive and more complex than InGaAs-based SPDs will be largely offset in star-type quantum networks, where a single BSM (i.e. two SNSPDs) can serve a large number of users. Next development steps of our system concern the integration of true random number generators, testing over real-world links exceeding distances of 100 km and in networks, and interfacing with quantum repeaters \cite{Sinclair14}.

\section*{Acknowledgements}
We acknowledge technical support by Vladimir Kiselyov, discussions with Erhan Saglamyurek and Neil Sinclair, and thank idQuantique for lending us two id210 single photon detectors. This work was supported through Alberta Innovates Technology Futures, the National Science and Engineering Research Council of Canada (through their Discover Grant and CryptoWorks 21 CREATE programs), the Calgary Urban Alliance, the National Nature Science Foundation of China (No. 61405030), the Oversea Academic Training Fund of the University of Electronic Science and Technology of China, the US Defense Advanced Research Projects Agency InPho Program, and the Killam Trusts. Part of the research was carried out at the Jet Propulsion Laboratory, California Institute of Technology, under a contract with the National Aeronautics and Space Administration. W.T. is a senior fellow of the Canadian Institute for Advanced Research.


\begin{thebibliography}{23}
\bibitem{BB84}
C.~H.~Bennett and G.~Brassard, {\em in Proc. IEEE Int. Conf. Comp. Sys. Sig. Process.} (Bangalore, 1984) pp. 175Ð179.

\bibitem{Gisin2002}
N.~Gisin, G.~Ribordy, W.~Tittel and H.~Zbinden, {\em Rev. Mod. Phys.} {\bf 2002}, {\em 74}, 145.

\bibitem{Scarani2009}
V.~Scarani, H.~Bechmann-Pasquinucci, N.~Cerf, M.~Du\u{s}ek, N.~L\"utkenhaus and M.~Peev, {\em Rev. Mod. Phys.} {\bf 2009}, {\em 81}, 1301.

\bibitem{Lo2014}
H.-K.~Lo, M.~Curty and K.~Tamaki, {\em Nat. Photon.} {\bf 2014}, {\em 8}, 595.

\bibitem{Stucki2009}
D.~Stucki, N.~Walenta, F.~Vannel, R.~T.~Thew, N.~Gisin, H.~Zbinden, S.~Gray, C.~R.~Towery and S.~Ten, {\em New J. Phys.} {\bf 2009}, {\em 11}, 075003.

\bibitem{Dixon2010}
A.~R.~Dixon, Z.~L.~Yuan, J.~F.~Dynes, A.~W.~Sharpe and A.~J.~Shields, {\em App. Phys. Lett.} {\bf 2010}, {\em 96}, 161102.

\bibitem{Sasaki2011}
M.~Sasaki, M.~Fujiwara, H.~Ishizuka, W.~Klaus, K.~Wakui, M.~Takeoka, S.~Miki, T.~Yamashita, Z.~Wang,  A.~Tanaka, K.~Yoshino, Y.~Nambu, S.~Takahashi, A.~Tajima, A.~Tomita, T.~Domeki, T.~Hasegawa, Y.~Sakai, H.~Kobayashi, T.~Asai, K.~Shimizu, T.~Tokura, T.~Tsurumaru, M.~Matsui, T.~Honjo, K.~Tamaki, H.~Takesue, Y.~Tokura, J.~F.~Dynes, A.~R.~Dixon, A.~W.~Sharpe, Z.~L.~Yuan, A.~J.~Shields, S.~Uchikoga, M.~Legr\'e, S.~Robyr, P.~Trinkler, L.~Monat, J.-B.~Page, G.~Ribordy, A.~Poppe, A.~Allacher, O.~Maurhart, T.~L\"anger, M.~Peev, and A.~Zeilinger, {\em Opt. Express} {\bf 2011}, {\em 19}, 10387.

\bibitem{Shields2013}
B.~Fr\"ohlich, J.~F.~Dynes, M.~Lucamarini, A.~W.~Sharpe, Z.~Yuan and A.~J.~Shields, {\em Nature} {\bf 2013}, {\em 501}, 69.

\bibitem{commercial}
See http://www.idquantique.com, http://www.sequrenet.com, http://www.quintessencelabs.com, http://www.quantum-info.com.

\bibitem{Brassard2000}
G.~Brassard, N.~L\"utkenhaus, T.~Mor and B.~C.~Sanders, {\em Phys. Rev. Lett.} {\bf 2000}, {\em 85}, 1330.

\bibitem{Gisin2006}
N.~Gisin, S.~Fasel, B.~Kraus, H.~Zbinden and G.~Ribordy, {\em Phys. Rev. A} {\bf 2006}, {\em 73}, 022320.

\bibitem{Makarov2006}
V.~Makarov, A.~Anisimov and J.~Skaar, {\em Phys. Rev. A} {\bf 2006}, {\em 74}, 022313.

\bibitem{LamasLinares2007}
A.~Lamas-Linares and C.~Kurtsiefer, {\em Opt. Express} {\bf 2007}, {\em 15} (15), 9388-9393.

\bibitem{Fung2007}
C.-H.~F.~Fung, B.~Qi, K.~Tamaki and H.-K.~Lo, {\em Phys. Rev. A} {\bf 2007}, {\em 75}, 032314.

\bibitem{Qi2007}
B.~Qi, C.-H.~F.~Fung, H.-K.~Lo and X.~Ma, {\em Quantum Inf. Comput.} {\bf 2007}, {\em 7}, 073.

\bibitem{Zhao2008}
Y.~Zhao, C.-H.~F.~Fung, B.~Qi, C.~Chen and H.-K.~Lo, {\em Phys. Rev. A} {\bf 2008}, {\em 78}, 042333.

\bibitem{Lydersen2010}
L.~Lydersen, C.~Wiechers, C.~Wittmann, D.~Elser, J.~Skaar and V.~Makarov, {\em Nat. Photon.} {\bf 2010}, {\em 4}, 686.

\bibitem{Jain2011}
N.~Jain, C.~Wittmann,  L.~Lydersen, C.~Wiechers, D.~Elser, C.~Marquardt, V.~Makarov and G.~Leuchs, {\em Phys. Rev. Lett.} {\bf 2011}, {\em 107}, 110501.

\bibitem{Tang2013}
Y.-L.~Tang, H.-L.~Yin, X.~Ma, C.-H.~F.~Fung, Y.~Liu, H.-L.~Yong, T.-Y.~Chen, C.-Z.~Peng, Z.-B.~Chen, and J.-W.~Pan, {\em Phys. Rev. A} {\bf 2013}, {\em 88}, 022308.

\bibitem{Bugge2014}
A.~N.~Bugge, S.~Sauge, A.~M.~M.~Ghazali, J.~Skaar, L.~Lydersen and V.~Makarov, {\em Phys. Rev. Lett.} {\bf 2014}, {\em 112}, 070503.

\bibitem{Dusek2000}
M.~Du\u{s}ek, M.~Jahma and N.~L\"utkenhaus, {\em Phys. Rev. A}  {\bf 2000}, {\em 62} , 022306. 

\bibitem{Yuan2010}
Z.~Yuan, J.~Dynes and A.~Shields, {\em Nat. Photon.} {\bf 2010}, {\em 4}, 800.

\bibitem{Lim2014}
C.~C.~W.~Lim, N.~Walenta, M.~Legr\'e, N.~Gisin and H.~Zbinden, {\em arXiv:1408.6398} {\bf 2014}.

\bibitem{Acin2007}
A.~Ac\`in, N.~Brunner, N.~Gisin, S.~Massar, S.~Pironio and V.~Scarani, {\em Phys. Rev. Lett.} {\bf 2007}, {\em 98}, 230501.

\bibitem{Masanes2011}
L.~Masanes, S.~Pironio and A.~Ac\`in, {\em Nat. Comm.} {\bf 2011}, {\em 2}, 238.

\bibitem{Giustina2013}
M.~Giustina, A.~Mech, S.~Ramelow, B.~Wittmann, J.~Kofler, J.~Beyer, A.~Lita, B.~Calkins, T.~Gerrits, S.~W.~Nam, R.~Ursin and A.~Zeilinger, {\em Nature} {\bf 2013}, {\em 497}, 227.

\bibitem{Christensen2013}
B.~G.~Christensen, K.~T.~McCusker, J.~B.~Altepeter, B.~Calkins, T.~Gerrits, A.~E.~Lita, A.~Miller, L.~K.~Shalm, Y.~Zhang, S.~W.~Nam, N.~Brunner, C.~C.~W.~Lim, N.~Gisin and P.~G.~Kwiat, {\em Phys. Rev. Lett.} {\bf 2013}, {\em 111}, 130406.

\bibitem{Lo2012}
H.-K.~Lo, M. Curty and B.~Qi, {\em Phys. Rev. Lett.} {\bf 2012}, {\em 108}, 130503.

\bibitem{Braunstein2012}
S.~L.~Braunstein and S.~Pirandola, {\em Phys. Rev. Lett.} {\bf 2012}, {\em 108}, 130502.

\bibitem{Gonzalez2014}
P.~Gonzalez, L.~Rebon, T.~Ferreira da~Silva, M.~Figueroa, C.~Saavedra, M.~Curty, G.~Lima, G.~B.~Xavier and W.~A.~T.~Nogueira, {\em arXiv:1410.1422}  {\bf 2014}.

\bibitem{Cao2014}
W.-F.~Cao, Y.-Z.~Zhen, Y.-L.~Zheng, Z.-B.~Chen, N.-L.~Liu, K.~Chen and J.-W.~Pan, {\em arXiv:1410.2928}  {\bf 2014}.

\bibitem{Geneva2014}
C.~C.~W.~Lim, B.~Korzh, A.~Martin, F.~Bussi\'eres, R.~Thew and H.~Zbinden, {\em arXiv:1410.1850} {\bf 2014}.



\bibitem{Biham1996}
E.~Biham, B.~Huttner and T.~Mor, {\em Phys. Rev. A} {\bf 1996}, {\em 54}, 2651.

\bibitem{Inamori2002}
H.~Inamori, {\em Algorithmica} {\bf 2002}, {\em 34}, 340.



\bibitem{Sangouard2011}
N.~Sangouard, C.~Simon, H.~de~Riedmatten and N.~Gisin, {\em Rev. Mod. Phys.} {\bf 2011}, {\em 83}, 33.

\bibitem{Rubenok2013}
A.~Rubenok, J.~A.~Slater, P.~Chan, I.~Lucio-Martinez and W. Tittel, {\em Phys. Rev. Lett.} {\bf 2013}, {\em 111}, 130501.

\bibitem{Liu2013}
Y.~Liu, T.-Y.~Chen, L.-J.~Wang, H.~Liang, G.-L.~Shentu, J.~Wang, K.~Cui, H.-L.~Yin, N-L.~Liu, L.~Li, X.~Ma, J.~S.~Pelc, M.~M.~Fejer, C.-Z.~Peng, Q.~Zhang and J.~W.~Pan, {\em Phys. Rev. Lett.} {\bf 2013}, {\em 111}, 130502.

\bibitem{daSilva2013}
T.~F.~da Silva, D.~Vitoreti, G.~B.~Xavier, G.~C. do~Amaral, G.~P.~Tempor\~ao and J.~P.~von der Weid, {\em Phys. Rev. A} {\bf 2013}, {\em 88}, 052303.

\bibitem{Tang2014}
Z.~Tang, Z.~Liao, F.~Xu, B.~Qi, L.~Qian and H.-K.~Lo, {\em Phys. Rev. Lett.} {\bf 2014}, {\em 112}, 190503.

\bibitem{Tang200}
Y.-L.~Tang, H.-L.~Yin, S.-J.~Chen, Y.~Liu, W.-J.~Zhang, X.~Jiang, L.~Zhang, J.~Wang, L.-X.~You, J.-Y.~Guan, D.-X.~Yang, Z.~Wang, H.~Liang, Z.~Zhang, N.~Zhou, X.~Ma, T.-Y.~Chen, Q.~Zhang and J.-W.~Pan, {\em Phys. Rev. Lett.} {\bf 2014}, {\em 113}, 190501.


\bibitem{Tang30}
Y.-L.~Tang, H.-L.~Yin, S.-J.~Chen, Y.~Liu, W.-J.~Zhang, X.~Jiang, L.~Zhang, J.~Wang, L.-X.~You, J.-Y.~Guan, D.-X.~Yang, Z.~Wang, H.~Liang, Z.~Zhang, N.~Zhou, X.~Ma, T.-Y.~Chen, Q.~Zhang and J.-W.~Pan, {\em IEEE Journal of Selected Topics in Quantum Electronics}  {\bf 2014}, {\em 21}, 6600407.

\bibitem{Curty2014}
M.~Curty, F.~Xu, W.~Cui, C.~C.~W.~Lim, K.~Tamaki and H.-K.~Lo, {\em Nat. Comm.} {\bf 2014}, {\em 5}, 3732.

\bibitem{Wang2013}
X.-B.~Wang, {\em Phys. Rev. A} {\bf 2013}, {\em 87}, 012320.

\bibitem{Ma2012}
X.~Ma, C.-H.~F.~Fung and M.~Razavi, {\em Phys. Rev. A} {\bf 2012}, {\em 86}, 052305.

\bibitem{Hwang2003}
W.-Y.~Hwang {\em Phys. Rev. Lett.} {\bf 2003}, {\em 91}, 057901.

\bibitem{Lodecoy}
H.~K.~Lo, X.~Ma and K.~Chen {\em Phys. Rev. Lett.} {\bf 2005} {\em 94}, 230504.

\bibitem{Wang2005}
X.-B.~Wang, {\em Phys. Rev. Lett.} {\bf 2005}, {\em 94} , 230503.

\bibitem{Xu2014}
F.~Xu, H.~Xu and H.-K.~Lo, {\em Phys. Rev. A} {\bf 2014}, {\em 89}, 052333.

\bibitem{Chan13}
P. Chan, J. A. Slater, I. Lucio-Martinez, A. Rubenok and W. Tittel, {\em Opt. Express}  {\bf 2014}, {\em 22}, 12716.

\bibitem{BSM50}
N.~L\"utkenhaus, J.~Calsamiglia and K.~A.~Suominen, {\em Phys. Rev. A} {\bf 1999}, {\em 59}, 3295.

\bibitem{Raju2014} 
R. Valivarthi, I. Lucio-Martinez, A. Rubenok, P. Chan, F. Marsili, V. B. Verma, M. D. Shaw, J. A. Stern, J. A. Slater, D. Oblak, S. W. Nam and W. Tittel, {\em Opt. Express}  {\bf 2014}, {\em 22}, 24497.

\bibitem{snspds}
F.~Marsili, V.~B.~Verma, J.~A.~Stern, S.~Harrington, A.~E.~Lita, T.~Gerrits and S.~W.~Nam, {\em Nat. Phot.} {\bf 2013}, {\em 7}, 210.

\bibitem{Corning}
http://www.corning.com

\bibitem{application note}
http://www.idquantique.com/photon-counting/support/resource-center-for-infrared-photon-counters.html

\bibitem{Sinclair14}
N.~Sinclair, E.~Saglamyurek, H.~Mallahzadeh, J.~A.~Slater, M.~George, R.~Ricken,M.~P.~Hedges, D.~Oblak, C.~Simon, W.~Sohler and W.~Tittel, {\em Phys. Rev. Lett.} {\bf 2014}, {\em 113}, 053603.

\bibitem{Gabriel10}
C. Gabriel, C. Wittmann, D. Sych, R. Dong, W. Mauerer, U. L. Andersen, C. Marquardt and G. Leuchs, {\em Nat. Photon.} {\bf 2010}, {\em 4}, 711.

\bibitem{LiuGuo2013}
Y. Liu, M.-Y. Zhu, B. Luo, J.-W. Zhang and H. Guo, {\em Laser Phys. Lett.} {\bf 2013}, {\em 10}, 045001.

\bibitem{England14}
D. G. England, P. J. Bustard, D. J. Moffatt, J. Nunn, R. Lausten and B. J. Sussman, {\em App. Phys. Lett.} {\bf 2014}, {\em 104}, 051117.



\end{thebibliography}
\end{document}